\documentclass[fleqn,usenatbib]{mnras}

\usepackage{newtxtext,newtxmath}

\usepackage[T1]{fontenc}
\usepackage{ae,aecompl}
\usepackage{comment}

\usepackage{graphicx}	
\usepackage[export]{adjustbox}

\usepackage{color}

\newcommand{\psr}{{PSR J0538+2817}}

\title[Radio filament of PSR~J0538+2817]{Discovery of a one-sided radio filament of PSR~J0538+2817 in S147:\\
escape of relativistic PWN leptons into surrounding supernova remnant?
}

\author[Khabibullin et al.]{I.I.~Khabibullin,$^{1,2,3}$ E.M.~Churazov,$^{2,3}$ A.M.~Bykov,$^4$ N.N.~Chugai,$^5$ and I.I.~Zinchenko$^{6}$
\\
\\
$^1$~Universitäts-Sternwarte, Fakultät für Physik, Ludwig-Maximilians-Universität München, Scheinerstr.1, 81679 München, Germany \\
$^2$~Max Planck Institute for Astrophysics, Karl-Schwarzschild-Str. 1, D-85741 Garching, Germany  \\
$^3$~Space Research Institute (IKI), Profsoyuznaya 84/32, Moscow 117997, Russia \\
$^4$~Ioffe Institute, 26 Politekhnicheskaya str., St. Petersburg 194021, Russia \\
$^5$~Institute of Astronomy, Russian Academy of Sciences, 48 Pyatnitskaya str., Moscow 119017, Russia  \\
$^6$~Institute of Applied Physics of the Russian Academy of Sciences, 46 Ul'yanov~str., Nizhny Novgorod 603950, Russia. 
}

\begin{document}
\label{firstpage}
\pagerange{\pageref{firstpage}--\pageref{lastpage}}
\maketitle

\begin{abstract} 
{We report the discovery of a faint radio filament near PSR~J0538+2817 in the NVSS, CGPS, and the Rapid ASKAP Continuum Survey data. This pulsar is plausibly associated with the supernova that gave rise to the Spaghetti Nebula (Simeis 147).}
The structure {is one-sided and} appears to be almost aligned {(within 17 degrees)} with the direction of the pulsar's proper motion, but in contrast to the known cases of pulsar radio tails, it is located ahead of the pulsar.  
At the same time, this direction is also {approximately} (within 5 degrees) perpendicular to the axis of the extended non-thermal X-ray emission {around} the pulsar. No X-ray or optical emission is detected from the filament region, although the end point of the radio filament appears to be adjacent to a filament of H$_\alpha$ emission. We speculate that this structure might represent a filament connecting pulsar wind nebula with the  {ambient} interstellar medium filled with relativistic electrons escaping the pulsar nebula, i.e. a radio analogue of X-ray filaments of Guitar and Lighthouse PWNs and filaments of non-thermal radio emission in the Galactic Center. 


\end{abstract}





\section{Introduction}

The collapse of the stellar core of massive ($M\gtrsim8M_{\odot}$) stars results in the energetic shock wave being launched, which is capable of disrupting the star, accelerating the debris to large (>10,000 km/s) velocities, and giving rise to the spectacular supernova phenomenon \citep[e.g.][]{2012ARNPS..62..407J}. In many cases, the collapse of the core also leads to the formation of a rotating and magnetized neutron star, which are believed to form a diverse population of isolated neutron stars, magnetars and pulsars \citep[e.g.][]{2012Ap&SS.341..457P}.

Pulsars are capable of tapping a certain {fraction} of their rotational energy into flows of highly relativistic particles in the form of equatorial wind, which, after being stopped by the surrounding medium, form the so-called Pulsar Wind Nebula \citep[PWN, e.g.,][]{2006ARA&A..44...17G}. PWNe are considered as an important source of Galactic leptons of very high energies, but the exact way how this escape takes place is still unclear \citep[][]{2017SSRv..207..235B,2018MNRAS.480.5419B}.  

Since the lifetime of pulsars is long, many of them having high enough initial velocity manage to overrun the {decelerating} supernova forward shock wave and start propagating through the undisturbed ISM \citep[e.g., ][]{2006ApJ...643..332F}. Interaction of a PWN with the ISM gives rise to the bow shock, advancing ahead of the rapidly moving pulsars \citep[e.g., ][]{2014ApJ...784..154B}. 

{Besides that, the linear tail-like structures are observed in some cases, probably being remnants of the PWN stripped by ram pressure of the inflowing ISM.  For instance, the cases of the bow shock and 6'-7' radio tails were found in PSR~J0002+6216  \citep[the Cannonball Pulsar, ][]{2019ApJ...876L..17S,2023ApJ...945..129K}.}

Radio imaging and polarization observations of a bow-shock PWN produced by PSR  J1437-5959 \citep[at Molonglo Observatory Synthesis Telescope and ATCA, ][]{2012ApJ...746..105N} showed about 10' extension nearly linear filament in the Frying Pan (G315.9-0.0) supernova remnant directed nearly radially outward from the rim of the shell. The magnetic field geometry inferred from radio polarimetry shows a good alignment with the tail orientation, which could be a result of high flow speed. There are also hints that the postshock wind has a low magnetization and is dominated by electrons and positrons in energy. Also, a tail is seen in the fast-moving pulsar PSR J0908-4913 associated with SNR \citep{2021MNRAS.507L..41J}. Numerous radio-emitting filaments observed in the MeerKAT \citep{2022ApJ...925..165H,2022ApJ...925L..18Y} and Karl Jansky Very Large Array \citep{2022ApJ...941..123P} data of the Galactic Center might also be relics of the pulsar-injected particles \citep[e.g.,][]{2019MNRAS.489L..28B}. 

Even more {intriguing} are perpendicular linear structures observed in X-rays in the Guitar \citep[see e.g.][]{2007A&A...467.1209H,2010MNRAS.408.1216J,2022ApJ...939...70D} and Lighthouse nebulae \citep{2014A&A...562A.122P} {, as well as from PSR~J1509-5850 \citep[][]{2016ApJ...828...70K}}. Recently, \citet{2023ApJ...950..177K}  reported on \textit{NuSTAR} observations of PSR~J1101-6101 and its misaligned outflow (which is the Lighthouse nebula) and detected the outflow up to 25 keV. They find clear evidence of spectral X-ray cooling with distance from the pulsar. These might be a result of reconnection of the PWN magnetic field lines with the field lines of the ISM, resulting in a flow of very energetic particles escaping  {PWNe and producing elongated synchrotron X-ray filaments \citep[][]{2008A&A...490L...3B,2017SSRv..207..235B,2019MNRAS.485.2041B,2019MNRAS.490.3608O}.}

On the other hand, some of the extended non-thermal emission structures associated with pulsars, are observed inside supernova remnants. An elongated structure called Vela X cocoon which is bright in X-rays and gamma-rays is apparently located inside the Vela SNR and can be associated with Vela PWN \citep[][]{2018ApJ...865...86S}. From the X-ray spectral data analysis the authors found in the cocoon the likely presence of both shocked ejecta material and the non-thermal PWN emission components. The X-ray data can possibly be interpreted as the result of a disruption of the Vela PWN by the asymmetric reverse shock of the Vela SNR. The asymmetry of the reverse shock was attributed to the large scale density gradient as it was predicted by \citet{Blondin_Chevalier01}.  

Here we report the discovery of a radio-emitting filament pointing towards PSR~J0538+2817 associated with the Simeis~147 SNR \citep[S147, e.g.][]{1976AZh....53...38L}, commonly referred to as the Spaghetti Nebula. We argue that the pulsar is likely still located within the realm of its parent supernova remnant and might be interacting either with the unshocked ejecta, shocked ejecta, or shocked interstellar medium. Orientation of the filaments precludes its interpretation as a PWN's stripping tail, and allows us to put forward several possibilities for its formation, depending on the assumed 3D position of the PSR inside the nebula. We discuss them in relation to the previous works which considered different phases of the PWN-SNR interaction \citep{Blondin_Chevalier01,swaluw04,Blondin_Chevalier17,2023arXiv230902263O} and formulate predictions of possible scenarios for future observations.

\section{Radio observations}

\begin{figure*}
\centering
\includegraphics[angle=0,bb=40 190 550 650, width=0.99\columnwidth]{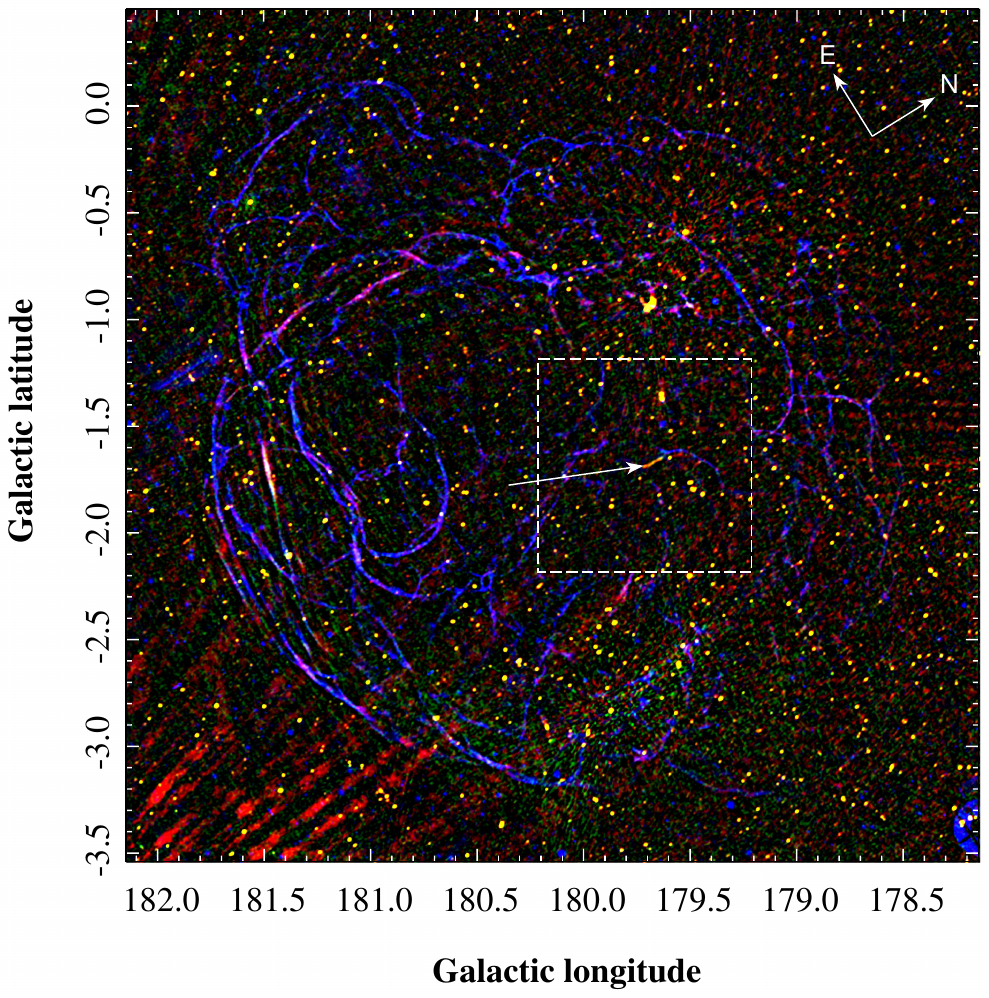}
\includegraphics[angle=0,bb=40 185 550 630, width=0.99\columnwidth]{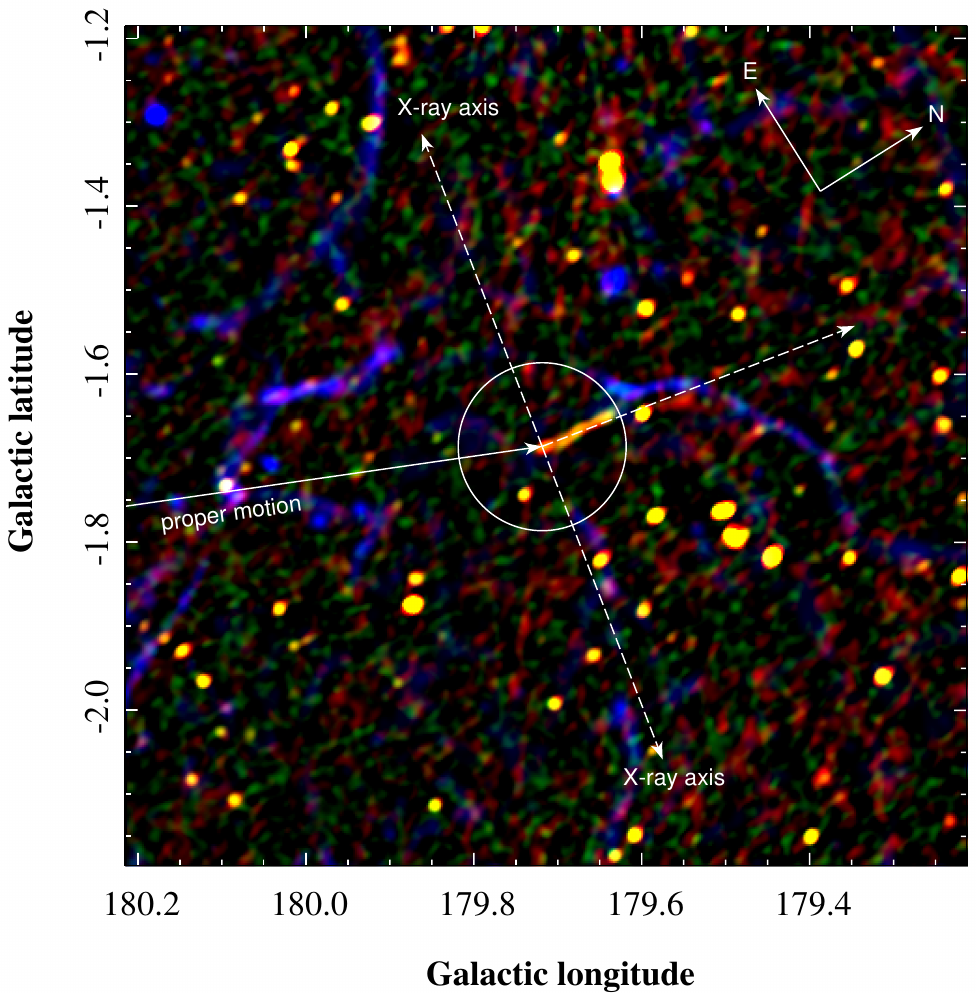}
\caption{Composite RGB images showing slightly smoothed RACS maps at 887.5 MHz (in red) and 1367.5 MHz (in green) and wavelet-filtered (to emphasise the filamentary structure) H$_{\alpha}$ image from the IGAPS survey (in blue). The left panel shows the full extent of the S147 nebula with the location of the PSR pointed by the arrow, which direction and length correspond to the proper motion of the pulsar for the last 40 kyrs. The square is 1 deg on a side and depicts the zoom-in region shown in the right panel. In addition to the direction of the proper motion, the right panel also shows the orientation of the mildly-extended X-ray emission detected by \textit{Chandra} above 4 keV, and the direction perpendicular to it. The circle is centred on the pulsar and has a radius of 6'. {The white compass region shows equatorial (J2000) North (N) and East (E) directions.}}
\label{fig:full}
\end{figure*}

Both  \psr~ \citep[][]{1996ApJ...468L..55A,2003ApJ...593L..31K,2007ApJ...654..487N,2021NatAs...5..788Y} and its host Spaghetti nebula \citep[][]{1974AJ.....79.1253D,1980PASJ...32....1S,1986A&A...163..185F,2008A&A...482..783X} have been extensively studied at radio frequencies, what allowed the pulsar period (134 ms), its derivative (corresponding to spin-down luminosity $5\times10^{34}$ erg/s), as well as its parallax \citep[corresponding to the distance of 1.4 kpc, e.g., ][]{2007ApJ...654..487N} and proper motion (corresponding to picture plane velocity 400 km/s in the direction away from the S147's center) being measured. Moreover, recent observations \citep[][]{2021NatAs...5..788Y} put a constrain on the line-of-sight velocity $\varv=81^{+158}_{-150}$ km/s, implying the full 3D velocity being smaller than 500 km/s (at 1$\sigma$ level). Assuming close-to-spherical shape of the main S147 shell, this also implies location of \psr\, well within its parent SNR boundaries. {3D alignment of the suggested pulsar rotation axis and its proper motion might be an indication of an explosion featuring electromagnetic rocket effect and launching of a powerful relativistic jet \citep[][]{2022MNRAS.509.4916X}.}

Here we report images of the S147 region based on the publicly available data of the Rapid ASKAP Continuum Survey (RACS) by Australian Square Kilometre Array Pathfinder (ASKAP) at 887.5 MHz \citep[RACS-low,][]{2020PASA...37...48M} and 1367.5 MHz \citep[RACS-mid,][]{2023PASA...40...34D}. Figure \ref{fig:full} shows an RGB composite images (in Galactic coordinates) combining slightly smoothed (to suppress noise) RACS maps at 887.5 MHz (in red) and 1367.5 MHz (in green) along with a wavelet-filtered (to emphasise the filamentary structure) H$_{\alpha}$ image from the IGAPS survey \citep[in blue][]{2021A&A...655A..49G}. Clearly visible in the left panel of Figure \ref{fig:full} is a close correspondence of some of the radio (primarily at lower frequencies) and H$_\alpha$ emitting filaments, while some of the bright H$_\alpha$ features lack radio counterparts. Although being of great interest as sites of energetic interaction between cold, hot and relativistic phases of the ISM, further consideration of these filaments is not the subject of the current paper.

Instead, a distinct filamentary structure which is visible (in the centre of the dashed rectangular 1x1 deg$^2$ region) both in 887.5 MHz and 1367.5 MHz images, lacks optical counterpart, as highlighted in the right panel of Figure \ref{fig:full}, and appears to be closely connected to the pulsar location is of interest here.

The length of this almost linear structure is $\approx6$ arcmin ($\sim2.5$ pc in projection at $d=1.4$ kpc), while its direction turns out to be aligned with the direction of the pulsar's proper motion (the motion of the pulsar over the last 40 kyrs is shown with the white arrow in Figure \ref{fig:full}, based on the proper motion measurements by \citealt{2007ApJ...654..487N}), but the filament is located in front (or ahead) of the pulsar, contrarily to the known radio tails of other pulsars.

At the same time, the direction of the filament appears to be as well close to the direction perpendicular to the elongation axis of the non-thermal X-ray emission from the PWN observed by \textit{Chandra} in the direct vicinity of the pulsar (this direction is shown via dashed line in the right panel of Figure \ref{fig:full}). In contrast to the H$_{\alpha}$ filaments of the Spaghetti nebula, which are also visible in (mostly low frequency) radio emission, the pulsar filament lacks optical counterpart (but see Section \ref{s:xray} for the discussion of possible connection of the filament's endpoint to the bright H$_{\alpha}$ substructure). No similar radio structures are visible along the way of pulsar from the center of S147. 

\begin{figure*}
\centering
\includegraphics[angle=0,bb=40 200 540 650, width=0.995\columnwidth]{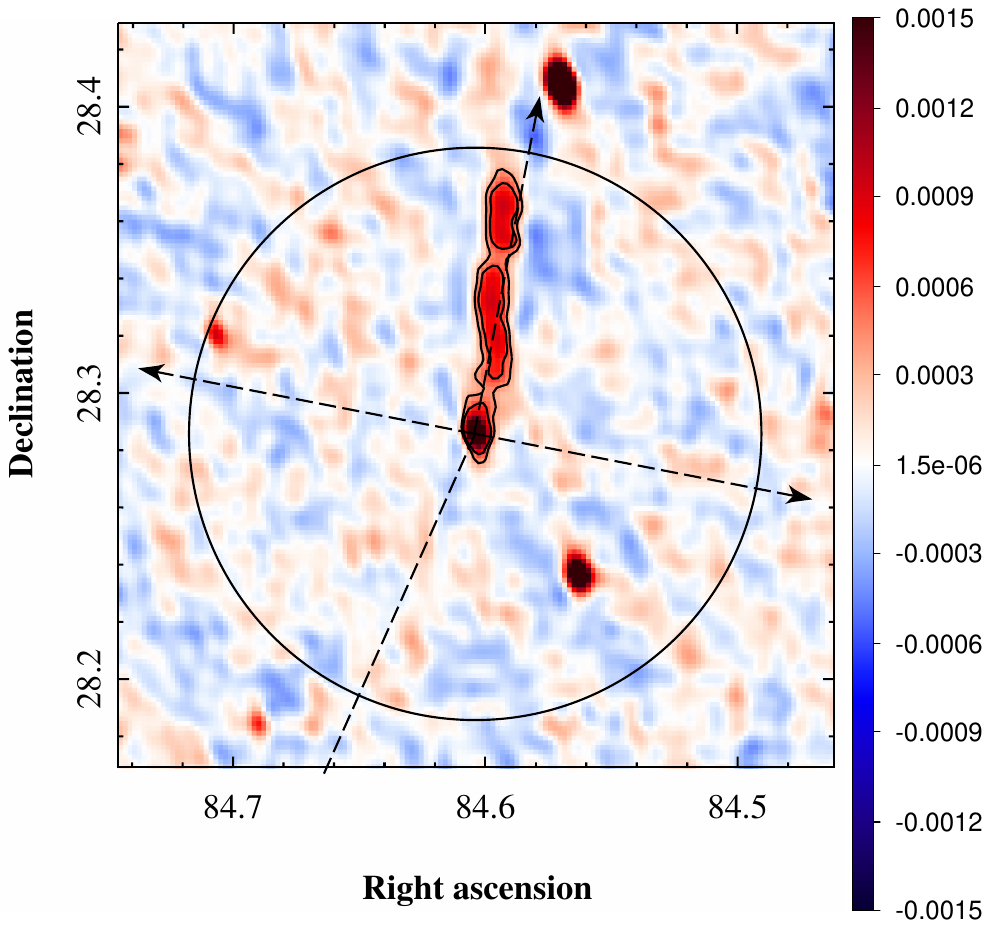}
\includegraphics[angle=0,bb=40 200 540 650, width=0.995\columnwidth]{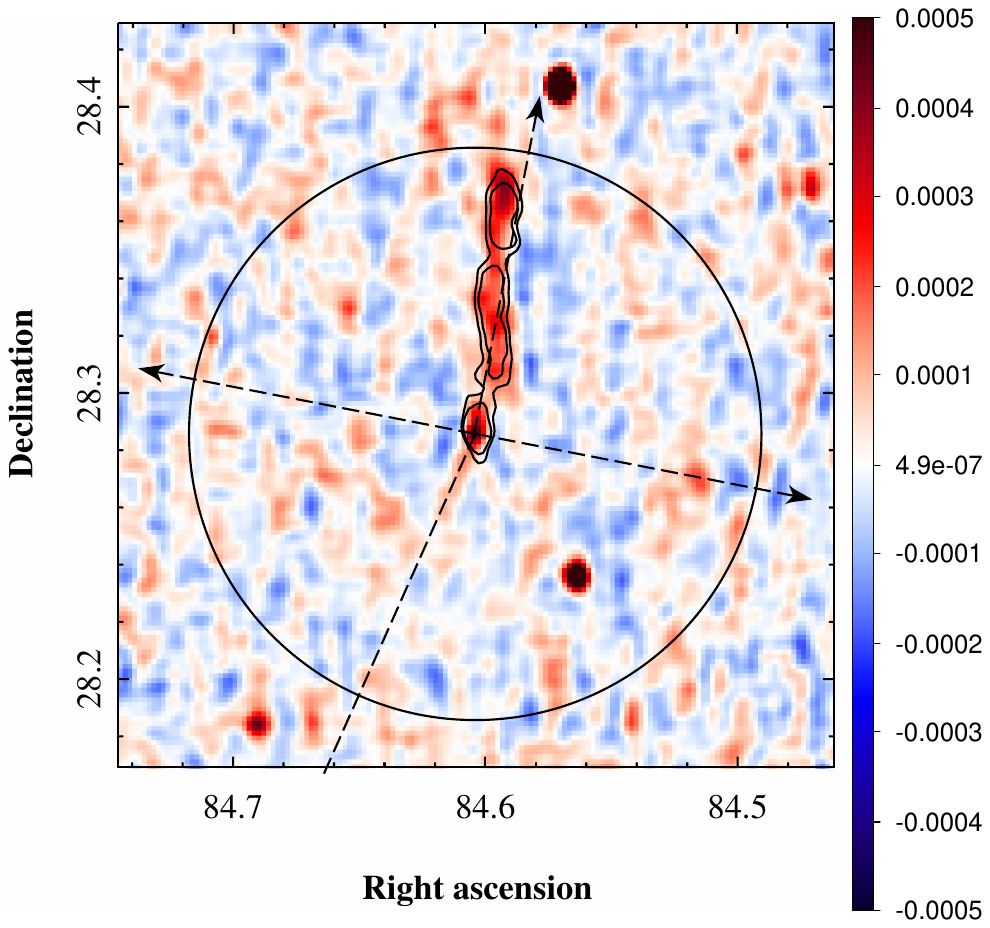}
\caption{Morphology of the radio emission {(in Jy/beam) at RACS low (887.5~MHz, left) and mid (1367.5~MHz, right)} frequencies versus the noise level and beam shapes, and in the equatorial coordinates (J2000) for easier comparison with previous studies. The black contours show the 3 and 5 RMS levels of the low-frequency map {($\sigma_{\rm low}\simeq135\mu$Jy/beam)} both on the left and right panels. Regions are as in Fig.~\ref{fig:full}, the circle is centred on the pulsar and has a radius of 6'.}
\label{fig:zoom}
\end{figure*}

Figure \ref{fig:zoom} shows morphology of the radio emission at 887.5 MHz (left panel) and 1367.5 MHz (right panel) in equatorial coordinates (J2000) and on the linear scale. The colour-scheme is set symmetrically around the zero level, so that the white colour corresponds to zero level, while blue regions demonstrate level noise fluctuations. Given the differences in the beam size in the spectral channels, no significant difference in the morphology between the bands is visible (black contours on both left and right panels of Figure \ref{fig:zoom} show 3 and 5 RMS levels on the low frequency map allowing comparison of the emission morphologies across the bands). The only exception might be presence of the Northern non-linear extension visible in the 1367.5 MHz map, which significance is rather low, however.

\begin{figure*}
\centering
\includegraphics[angle=0,bb=50 200 540 650, width=0.68\columnwidth]{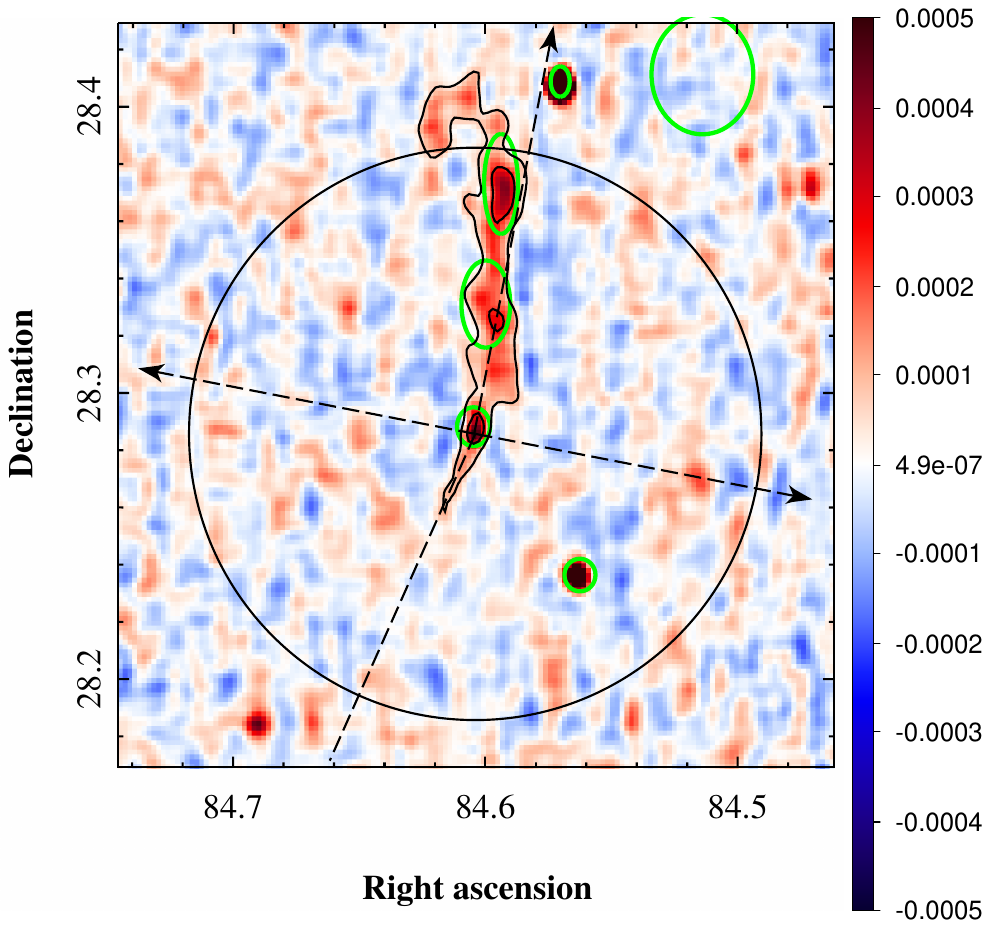}
\includegraphics[angle=0,bb=50 200 540 650, width=0.68\columnwidth]{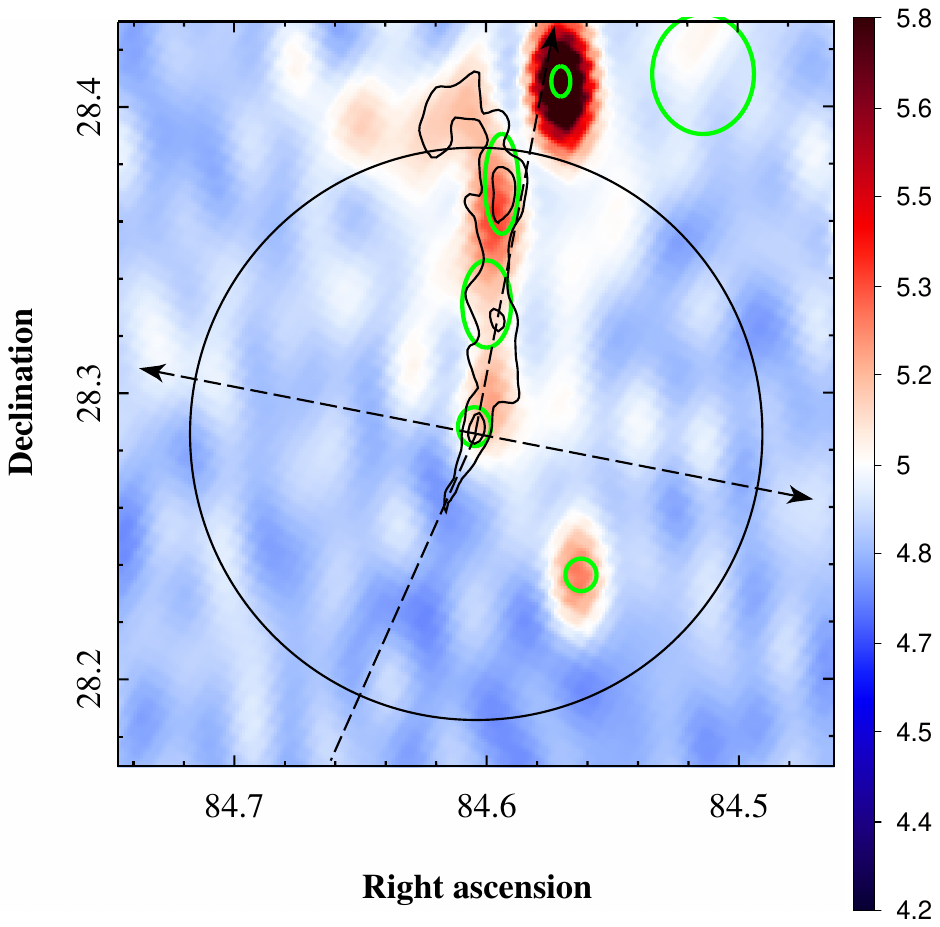}
\includegraphics[angle=0,bb=50 200 540 650, width=0.68\columnwidth]{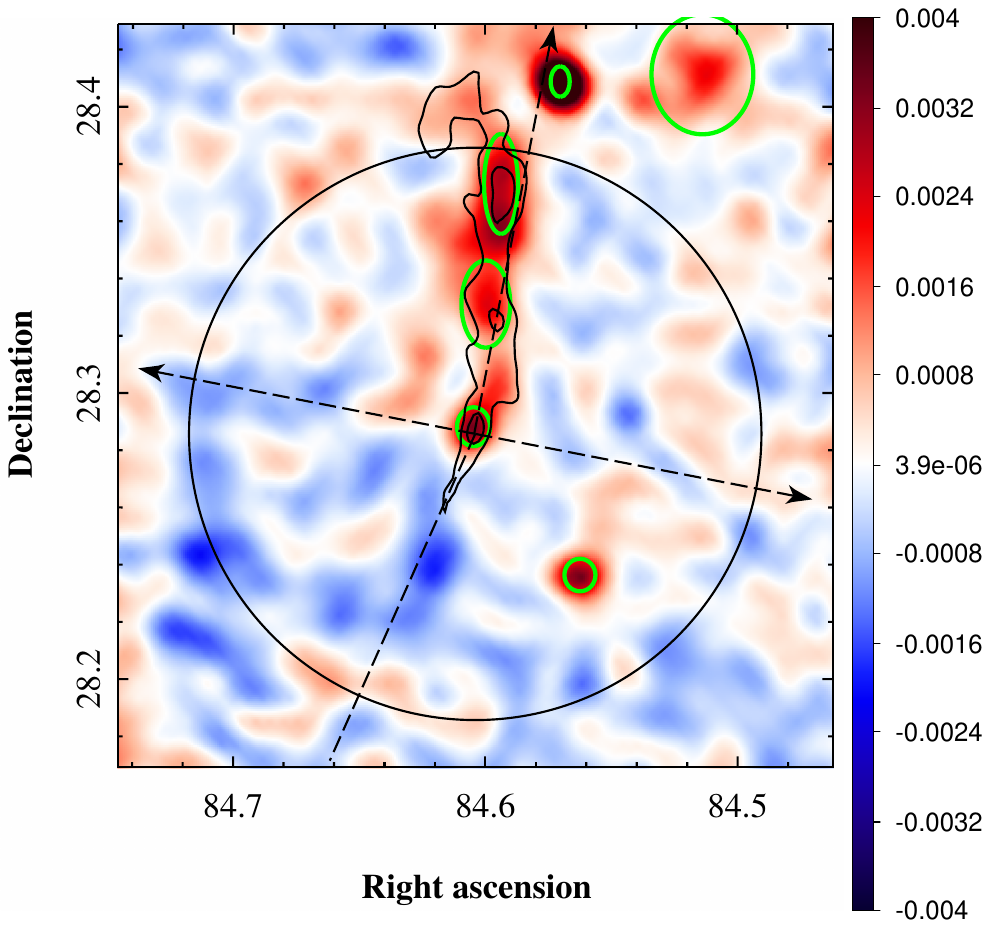}
\caption{Possible extension of the radio emission at 1.4 GHz based on the RACS-mid (left panel), CGPS (middle panel), and NVSS (right panel) data.
The contours correspond to the RACS-mid image, the other black regions are as in Figure \ref{fig:zoom}. Green regions show the locations and approximate shapes of the sources from the NVSS catalogue.}
\label{fig:nvsscgps}
\end{figure*}

Given that at low significance level, interferometric radio maps contain plenty of artefacts, in particular caused by bright point sources (cf. the bright red stripes in the bottom left part of Figure \ref{fig:full} caused by the Crab nebula), we also examine the archival data of the NVSS \citep[][]{1998AJ....115.1693C} and CGPS \citep{2003AJ....125.3145T} surveys at 1.4 GHz. Figure \ref{fig:nvsscgps} shows comparison the RACS-mid (left panel), CGPS (middle panel) and NVSS (right panel) images, where black contours reflect morphology of the filament emission in the ASKAP image and green regions mark positions and approximate extents of the sources in the NVSS catalogue \citep[][]{1998AJ....115.1693C}. Remarkably, the filament emission is visible in both NVSS and CGPS images, including the possible Northern extension. Moreover, this emission is recognized as a pair of mildly extended sources in the NVSS catalogue, with the combined flux density $\sim10$ mJy at 1.4 GHz, being $\sim3$ times brighter than the pulsar itself, having $3.5\pm0.5$ mJy at 1.4 GHz \citep[][]{1998AJ....115.1693C}. 

For \psr~, the flux density at 887.5 MHz is $7.5\pm1$ mJy at 888 MHz (ASKAP), so the the spectral index is close to $\sim 0.5$, and its radio luminosity is at the level of $\sim 1.5\times 10^{28}$ erg/s at $d=1.4$ kpc. The quality of the publicly available RACS data precludes us from drawing firm conclusions regarding the spectral shape of the diffuse filament emission, but simple "hardness ratio" comparison of the images indicates that it is not strongly dissimilar to the pulsar itself. As a result, we estimate $\nu L_{\nu}$ luminosity of the diffuse emission at the level of at least $\sim 5 \times 10^{28}$ erg/s, which is six order of magnitude below the pulsars spin-down luminosity  $L_{\rm sd}\sim 5 \times 10^{34}$ erg/s.

More robust radio data are needed to improve on these estimates, and the required sensitivity level is well within the reach of the currently operating facilities, so we leave a more elaborate analysis for future work. Here, we conclude that the radio filament emission is not dissimilar to the pulsars emission in spectral shape an luminosity and does not show strong variations between the epochs of radio observations separated by more than 20 yrs or so.

\begin{figure}
\centering
\includegraphics[clip=true, angle=0, trim=1.0cm 6.6cm 1.5cm 6cm,width=0.95\columnwidth]{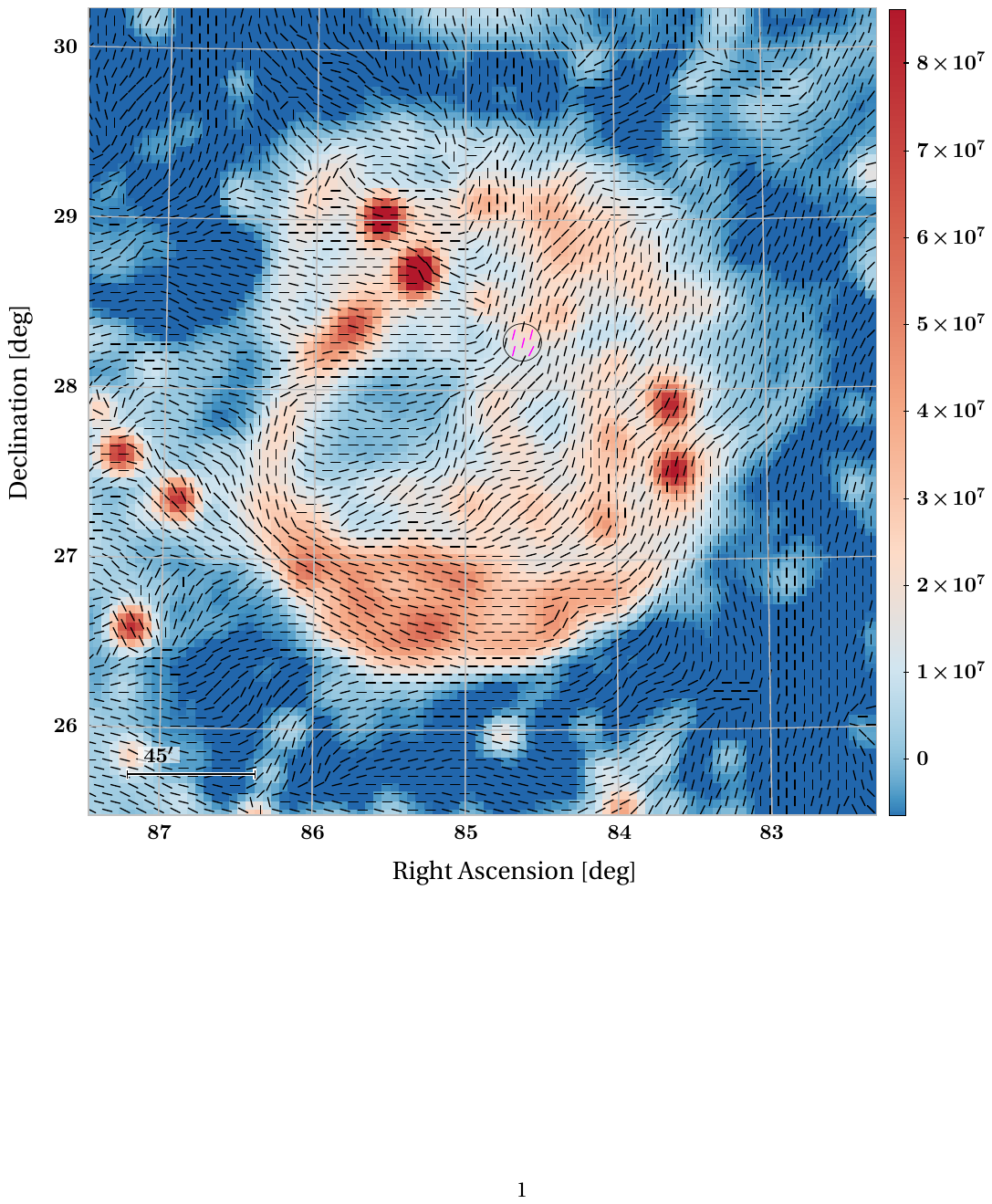}
\caption{Intensity and the magnetic field direction map (equatorial coordinates) derived from the Stokes parameters map at $\lambda=6$ cm by Urumqi telescope \citep[][]{2008A&A...482..783X}. An $r=6$ arcmin region around the pulsar is marked with the magenta colour, showing close-to-vertical orientation of the magnetic field, similarly to the orientation of the discovered radio filament.}
\label{fig:polarmap}
\end{figure}

Finally, we note that direction of discovered filament is also close to the magnetic field vector direction derived from the polarimetric observations of the Spaghetti nebula at $\lambda=6$cm (5 GHz) by Urumqi telescope \citep[][]{2008A&A...482..783X}. Figure \ref{fig:polarmap} shows the intensity map of the $\lambda=6$cm radio emission (linear scale, equatorial coordinates) with the direction of the magnetic field vector overlaid in black on top of it. Although, according with expectations in compression scenario, the magnetic field direction follows the brightest regions of the SNR's rim, no sharp change in the B direction across the SNR boundary is visible, indicating that the observed B direction is strongly affected, if not determined, by the B direction in the interstellar medium unaffected by the SNR shock wave \citep[][]{2008A&A...482..783X}. For the region of the Galaxy in direction of the S147 nebula, the magnetic field is known to aligned parallel to the Galactic disk, as demonstrated by the synchrotron and dust polarization maps obtained by \textit{Planck} \citep[e.g. Fig. 23 and 25 in ][]{2016A&A...594A...1P}. This direction, corresponding to vertical orientation in the equatorial coordinates is indeed very close to the orientation of the discovered radio feature. Given poor spatial resolution of the Urumqi map, however, it is difficult to draw any firm conclusions on the significance of this alignment and whether it would still hold when scales comparable to the size of the filament are resolved. Polarimetric observations with ASKAP or MeerKAT will certainly be invaluable for clarifying this.


\section{H$_\alpha$ view}
\label{s:halpha}

\begin{figure}
\centering
\includegraphics[angle=0,bb=40 160 530 680, width=0.99\columnwidth]{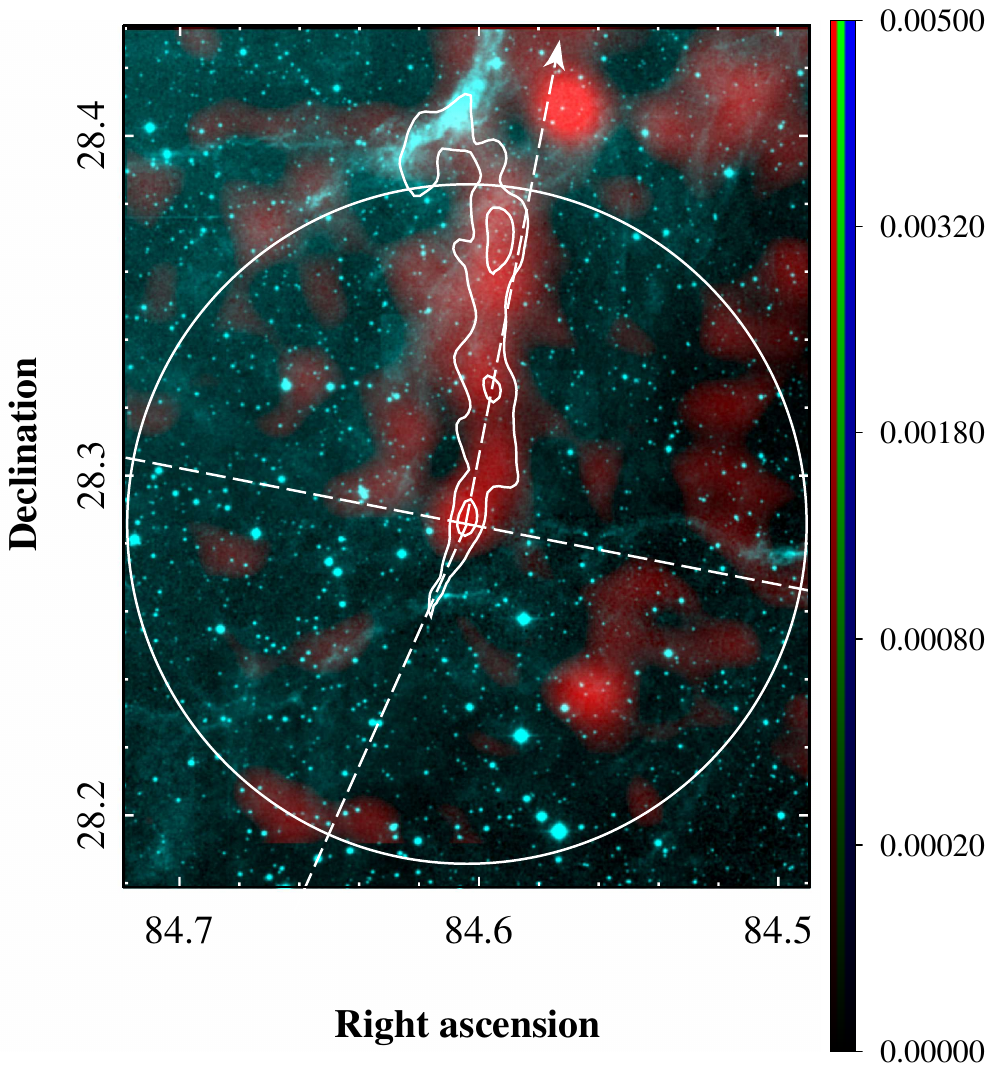}
\caption{Composite of H$_{\alpha}$ (IGAPS, cyan) and radio emission at 1.4 GHz (NVSS, red) with white contours showing morphology of the extended radio emission in the RACS 1367.5 MHz map. The circle is 6 arcmin in radius.}
\label{fig:nvssha}
\end{figure}

As was mentioned earlier, no optical counterpart of the radio structure is visible on deep IGAPS images as illustrated more explicitly on the composite image shown in Figure \ref{fig:nvssha}, combining NVSS image at 1.4 GHz in red and IGAPS H$\alpha$ image in cyan (equatorial coordinates) with the contours of 1367.5 MHz from RACS-mid overlaid in white. Clearly, the main straight body of the radio filament corresponds to a slight depression of the H$_\alpha$ emission, if anything, which is likely simply a result of fluctuations in the surface brightness of the emission from the Spaghetti nebula. 

The Northern extension of the radio filament visible at lower significance appears to coincide with the bright portion of an H$_{\alpha}$ filament, which might be either an indication of its unrelated (with respect to the pulsar) nature, or reveal a connection between the two structures.   

\begin{figure}
\centering
\includegraphics[angle=0,bb=50 220 540 640, width=1.00\columnwidth]{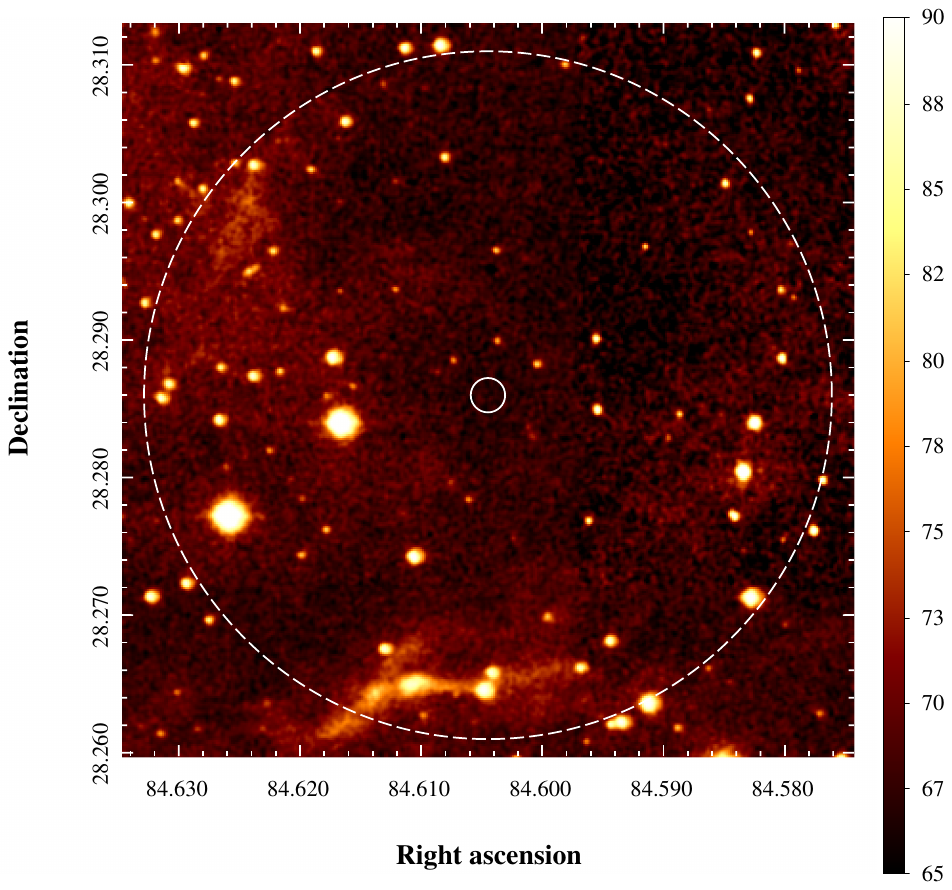}
\caption{H$_{\alpha}$ image of the \psr\, vicinity based on mosaic from IGAPS. The small circle is 4.5'' in radius, and the large circle is 20 times bigger. No signatures of the extended emission associated with a bow shock is visible at the level comparable to the diffuse emission from S147. }
\label{fig:halpha}
\end{figure}

The direct vicinity of the pulsar itself also appears to be dark in H$_\alpha$ emission (see Figure \ref{fig:halpha}), with no signatures of a PWN bow shock at least at the level corresponding to the H$_\alpha$ emission of the Spaghetti nebula, most likely arising from an SNR-driven shock propagation in the cold and neutral unperturbed interstellar medium. Thus, we conclude that \psr~ is indeed most likely still well within the boundary of its parent supernova remnant and propagates through a relatively tenuous unshocked ejecta or the hot interior between the forward and reverse shocks.


\section{X-ray view}
\label{s:xray}

\psr~ was a target of both \textit{Chandra} and \textit{XMM-Newton} observations, that on the one hand allowed to resolve the X-ray PWN \citep[e.g.][]{2003ApJ...585L..41R,2007ApJ...654..487N} and discover X-ray pulsations \citep[][]{2003ApJ...591..380M}. Here we {return to} these observations primarily in order to search for a possible X-ray counterpart of the newly discovered radio filament.

The initial processing of the \textit{Chandra} data was performed using the latest calibration data and following the standard procedure described in \citealt{2005ApJ...628..655V}. Corrections for the exposure map variations, the vignetting effect of the telescope and subtraction of the particle background were done {similarly} to \citealt{2012MNRAS.421.1123C}.
The ObsIDs used for the analysis here were 2796, 5538, 6242 (PI: Roger Romani).

Exquisite spatial resolution of \textit{Chandra} allows to resolve rich morphology of the X-ray emission at the arcsec scales around the pulsar. The soft X-ray emission extends away to $\sim7$ arcsec and has a wind-like shape. At higher energies,likely dominated by non-thermal emission of the PWN, the emission extends to 4.5 arcsec, and its axis is consistent with the measurement by \citealt{2007ApJ...654..487N}.

{\citealt{2007ApJ...654..487N} estimated the (unabsorbed) flux of the PWN emission at the level of $2.4\times10^{-14}$ erg/s/cm$^2$ (0.5-5 keV), resulting in the luminosity of $\sim 6.5 \times 10^{30}$ erg/s, i.e. $\eta\sim 10^{-4}$ of the spin-down luminosity.}

{
PSR J0538+2817 was observed with \textit{XMM-Newton} in 2002 (OBSID: 0112200401). Only one instrument (MOS1) of the European Photon Imaging Camera (EPIC) was operating in the imaging mode and covered the filament area. The other two detectors were in timing mode. We use MOS1 data for imaging and spectral analysis after subtraction of the particle and blank field background contributions and correction for exposure map variations and vignetting of the telescope. Figure \ref{fig:xmm} shows images (equatorial coordinates) of the X-ray surface brightness in 0.5-3 keV (left panel) and 3-7 keV (right panel) bands. 
No excess emission is visible from the filament region highlighted by the green contours taken from the map of radio emission at 887.5 MHz.
}

To obtain an upper limit on the total X-ray emission from the filament region, we define three box regions as indicated in cyan in Figure \ref{fig:xmm}, with the middle one fully encompassing the radio filament, and the other adjacent two used for the background estimation. As a result we get an upper limit on the 0.5-8 keV surface brightness at the level of $\sim 2\times10^{-15}$ erg/s/cm$^2$/arcmin$^2$ (which is $\sim10\%$ of the background level), resulting in the flux limit within the radio bright region at the level of  $\sim 10^{-14}$ erg/s/cm$^2$.

\smallskip
This translates into the luminosity limit at $d=1.4$ kpc $L_X\lesssim 3\times10^{30}$ erg/s. If the radio luminosity is at the level of $L_{\rm 1~GHz}\sim 4\times10^{28}$ erg/s at 1 GHz, this upper limit translates into a lower limit on the cross-band spectral index of $\alpha_{XR}\gtrsim0.75$.

On the other hand, we can put a lower limit on the magnetic field inside the filament, given that no X-ray emission coming from the Inverse Compton radiation of the same population of the electrons is observed at the level 100 times the radio luminosity (e.g. \citealt{1966ApJ...146..686F}, see also eq.5.10 in \citealt{1988xrec.book.....S}). 

For the spectral index of radio and X-ray emission $\alpha_{X}=\alpha_{R}=0.5$, this limit corresponds to $\sim40$ nG, conservatively assuming that the radiation field responsible for the IC comes primarily from the Cosmic Microwave Background radiation. For steeper spectra, this limit becomes even higher, corresponding to 300 nG for  $\alpha_{X}=\alpha_{R}=1.0$ {(cf. Fig.~\ref{fig:xrratio} showing dependence of the X-ray-to-radio ratio on the magnetic field strength and slope of the particle distribution function).} Although these values are still much smaller than one might expect for the (shocked) interstellar medium, in the unshocked ejecta case much smaller filed strengths might take place, given that there is no mechanism of the field amplification effectively operating in the free expansion phase.

\begin{figure*}
\centering
\includegraphics[angle=0,bb=40 190 550 670, width=0.9\columnwidth]{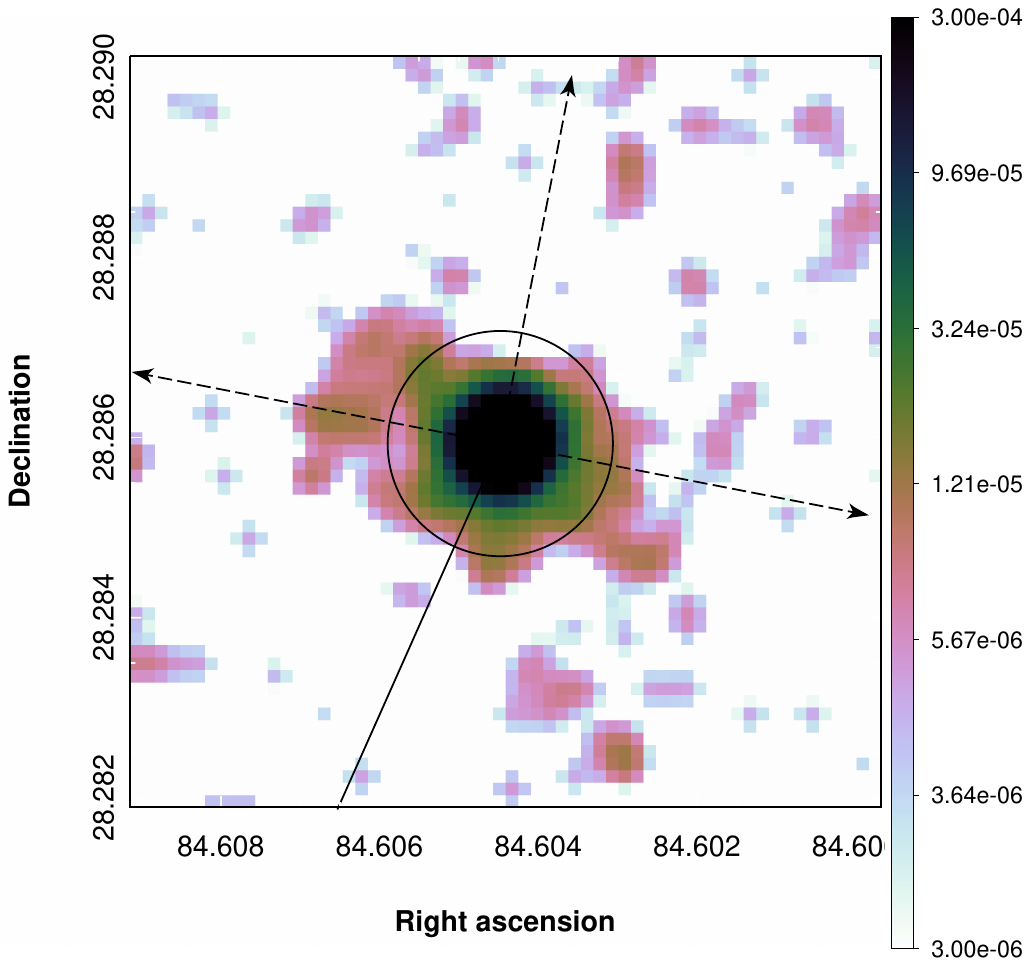}
\includegraphics[angle=0,bb=40 190 550 670, width=0.9\columnwidth]{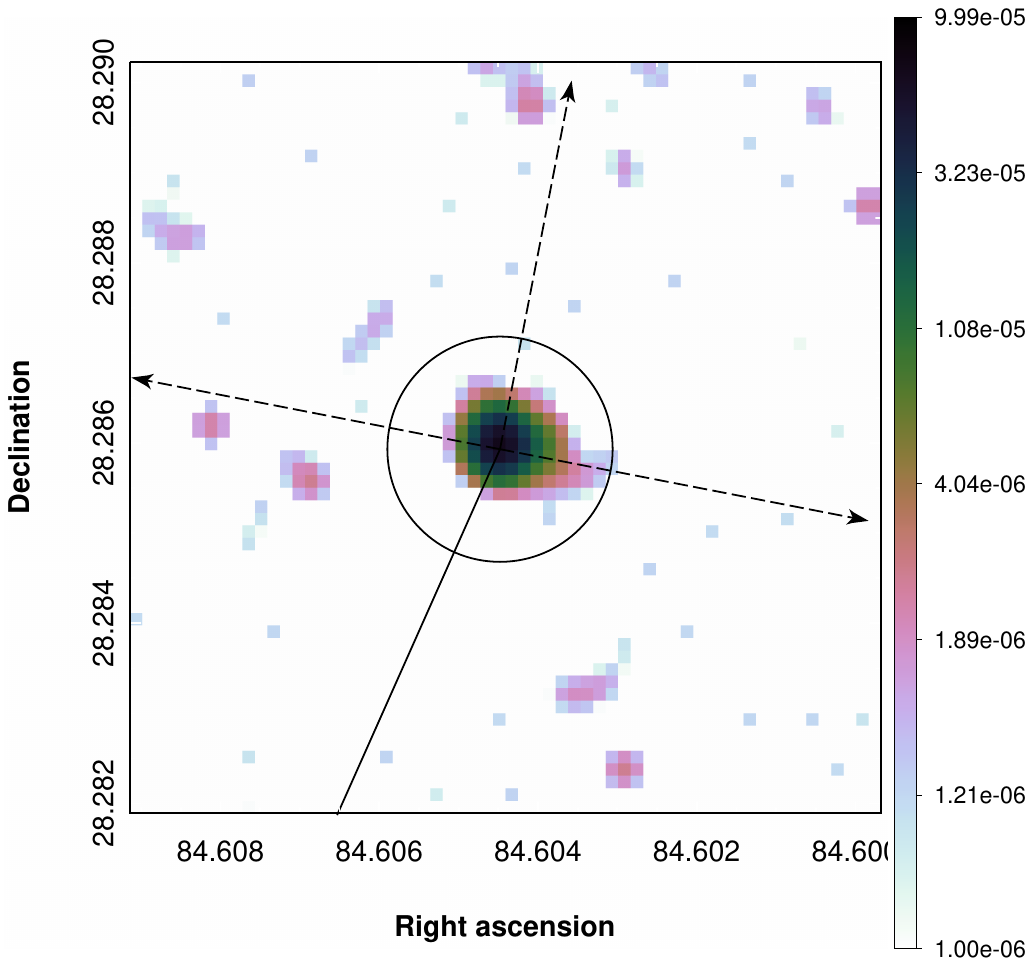}
\caption{\textit{Chandra} X-ray $0.5'\times0.5'$ images (equatorial coordinates) smoothed with $\sigma=0.5''$ Gaussian window in the 0.5-7 keV (top) and 3-7 keV (bottom bands) on a logarithmic scale. The solid circle is centred on the pulsar and has a radius of $4.5''$. The solid line shows the direction of the pulsar's proper motion, while the oppositely directed dashed arrows show the X-ray axis proposed by \citealt{2007ApJ...654..487N} and the perpendicular direction.}
\label{fig:chandra}
\end{figure*}

\begin{figure*}
\centering
\includegraphics[angle=0,bb=40 190 550 670, width=0.95\columnwidth]{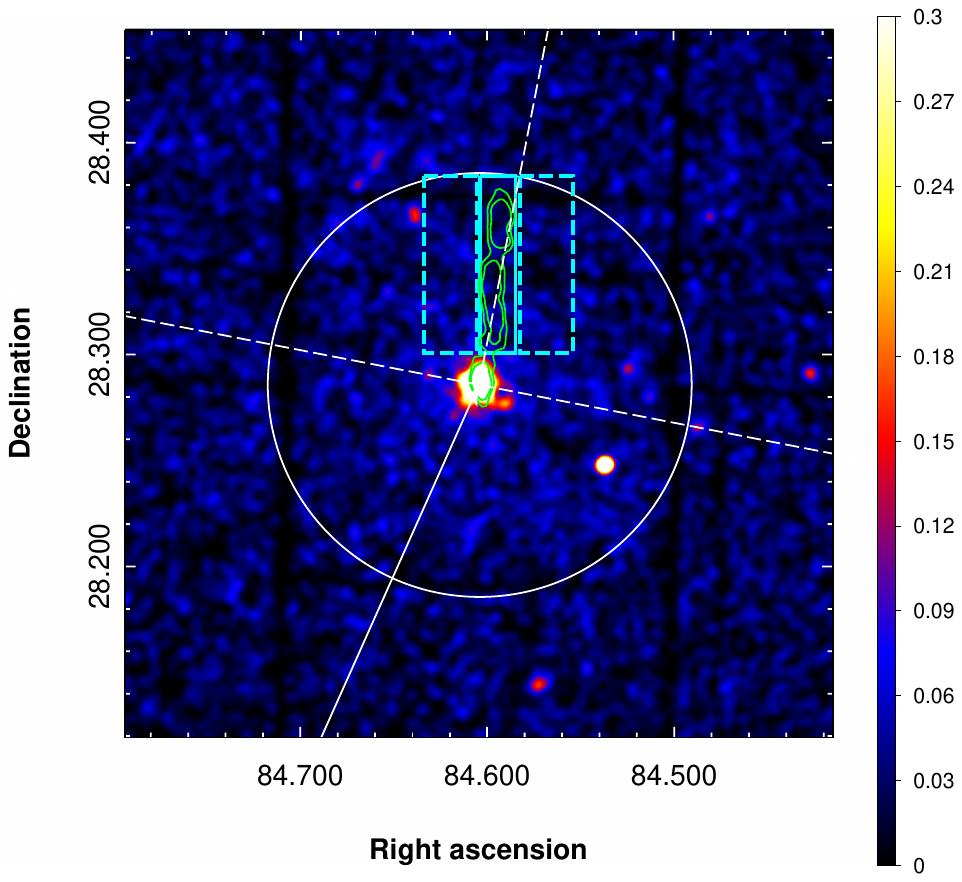}
\includegraphics[angle=0,bb=40 190 550 670, width=0.95\columnwidth]{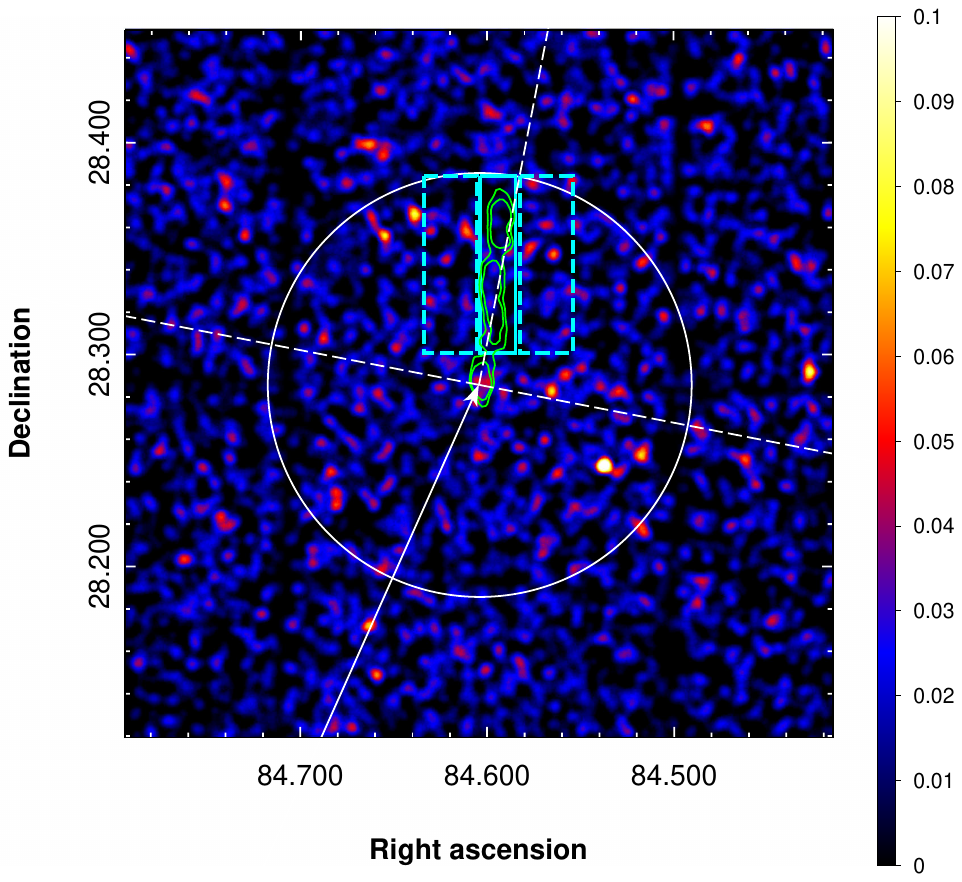}
\caption{\textit{XMM-Newton} images in the 0.5-3 keV (left panel) and 3-7 keV (right panel) bands (equatorial coordinates, linear scale) with contours of 887.5 MHz emission from the radio filament overlaid in green. So signatures of the excess X-ray emission from the location of the radio-emitting filament can be spotted neither in soft nor in hard X-rays. }
\label{fig:xmm}
\end{figure*}

\begin{figure}
\centering
\includegraphics[angle=0,bb=40 190 550 670, width=0.95\columnwidth]{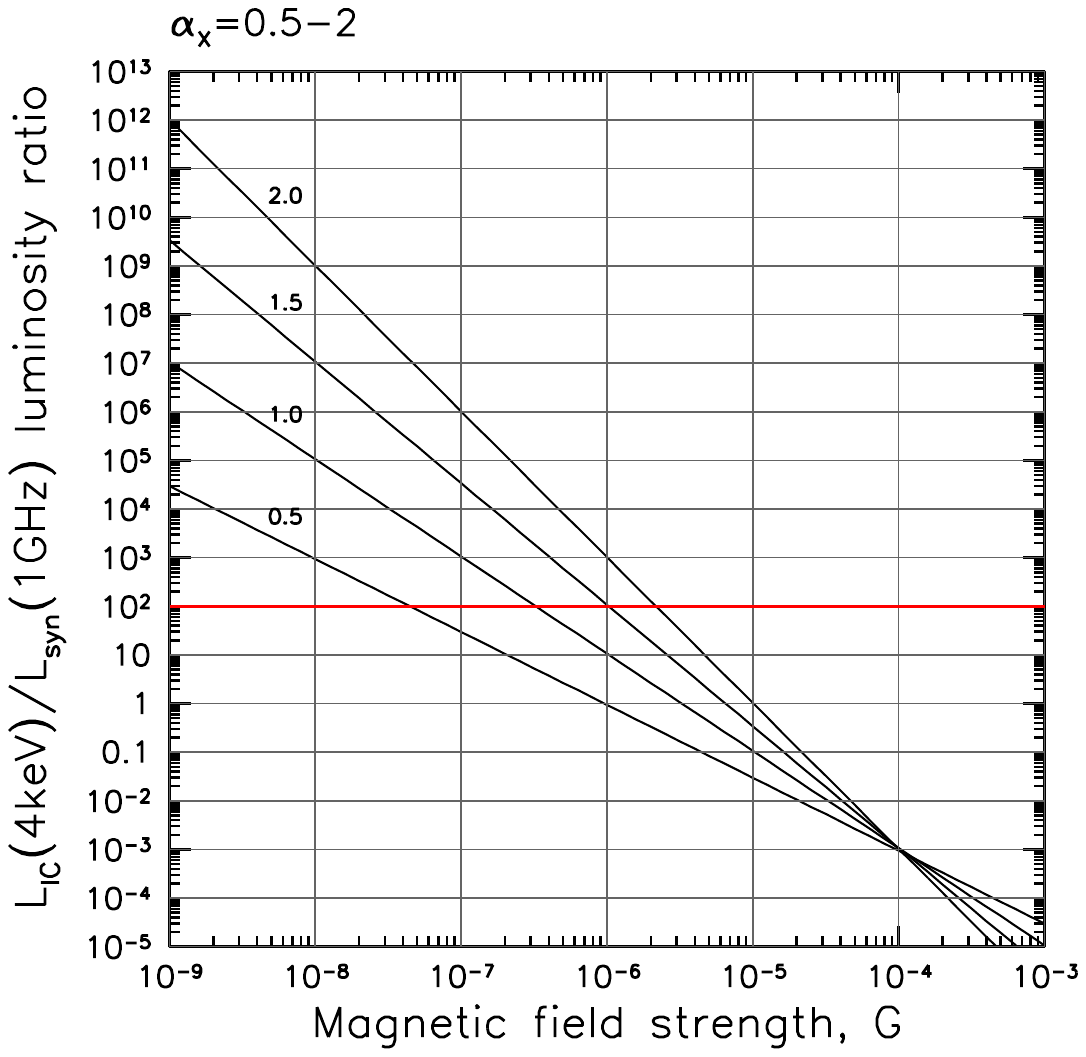}
\caption{ {Ratio of the X-ray (at 4 keV) and radio (at 1~GHz) $\nu L_{\nu}$ luminosities for a single population of relativistic particles emitting in Inverse Compton (from CMB radiation field) and synchrotron regimes, respectively.
The slope $p$ of the particle distribution function $dN/d\gamma \propto \gamma^{-p}$ is determined by the spectral index in X-ray band $\alpha_{X}=(p-1)/2$, which ranges from 0.5 to 2, as indicated next to each line. The red line marks the upper limit on this ratio obtained from the upper limit on the non-thermal X-ray emission from the region of the radio-emitting filament. For $\alpha_{X}\gtrsim0.5$, $B\gtrsim40$~nG is required.}}
\label{fig:xrratio}
\end{figure}
\section{Discussion}
\label{s:discussion}
\begin{figure*}
\centering
\includegraphics[angle=0, width=2.0\columnwidth]{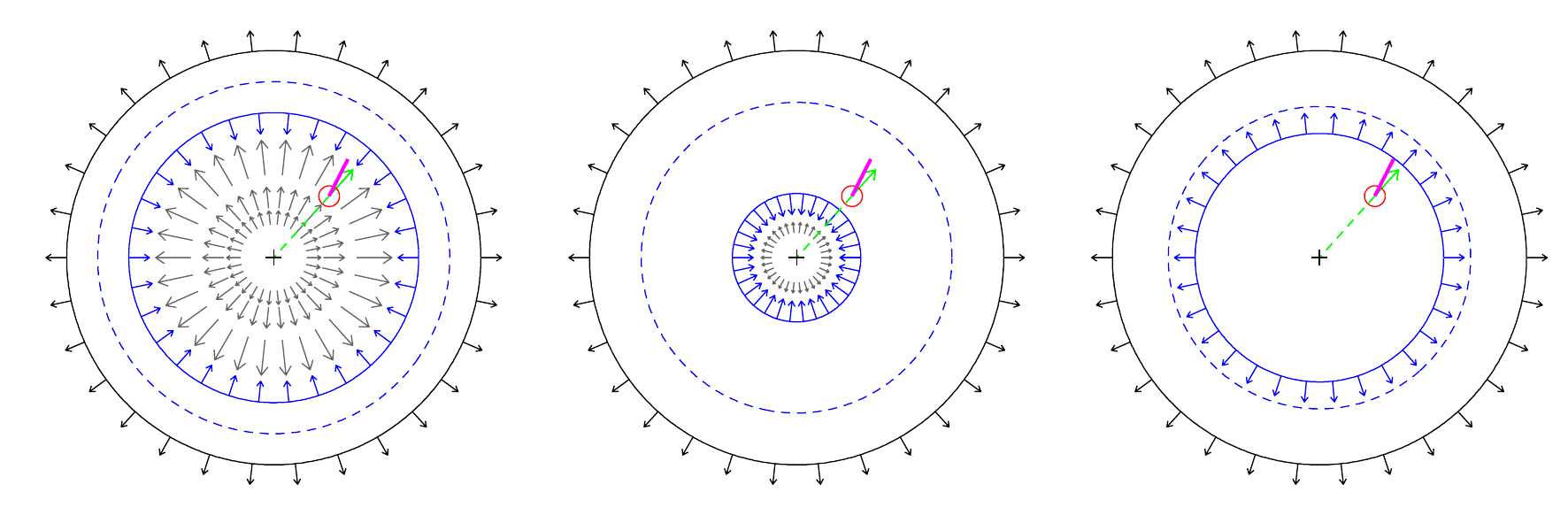}
\caption{Sketch of the three possible scenarios for location of the pulsar and its PWN (marked with the red circle and magenta line showing orientation of he radio filament) inside its parent supernova remnant (bounded by the outgoing spherically symmetric forward shock shown in black): left - pulsar is still propagating (direction of the proper motion is shown in green) along with the ejecta (which velocity field is schematically depicted in grey) yet unshocked by the reverse shock (blue); middle - the situation when the reverse shock has already passed the pulsar; right - the pulsar has been already passed by the outgoing reflected shock wave. Real geometry of the SNR shocks can differ strongly from the spherically symmetric one, resulting in more complicated flow patterns. }
\label{fig:sketch}
\end{figure*}

Let us first summarise the main observational findings regarding the newly discovered radio filament:
\begin{itemize}
    \item the radio filament is almost straight and narrow, with the length of $\sim$6 arcmin ($\sim2.5$ pc in projection at $d=1.4$ kpc) and at least $\sim10$ times smaller width
    \item {the filament is one-sided and, unlike ram pressure stripped tails, it points approximately {\it in the direction} of the PSR proper motion (within 17 degrees)}
    \item {filament's spectral shape and luminosity at GHz frequencies are not very dissimilar from the pulsar and its immediate vicinity, with the filament being a factor of 3 more luminous. Possible signs for spectral hardening can be seen towards the end of the filament and in particular in the Northern extension region. However, the latter could be an unrelated feature associated with ${\rm H_{\alpha}}$ filaments of the Spaghetti nebula }
    \item {comparison of the NVSS, CGPS, and ASKAP surveys does not reveal dramatic epoch-to-epoch variations on the time span of more than 20 years}
     \item {direction of the filament is also close to the perpendicular direction to the axis of X-ray emission of the PWN (within 5 deg). The filament is also aligned with the magnetic field direction derived from the low-resolution synchrotron polarization maps, although the latter could reflect global magnetic field orientation along the Galactic plane rather than inside S147}
    \item no X-ray or optical counterpart is detected, implying a lower limit on the magnetic field strength of $\sim40$ nG (for a power law distribution of electrons)
\end{itemize}

{At least four stages of the interaction of a pulsar and its PWN with the surrounding medium have been studied \citep[see e.g. a recent review by][and their Fig. 1]{2023arXiv230902263O}, broadly corresponding to the PSR position relative to the SNR shock waves. We briefly discuss below all four scenarios for the \psr\, and acknowledge related issues. 

These stages are 
\begin{enumerate}
\item PSR moving together with free-expanding ejecta
\item PSR passing through the reverse shock and moving through the shocked ejecta or shocked ISM
\item The reverse shock bounces from the center and overtakes outwardly moving PSR
\item PSR escapes from the SNR and moves through the ISM. 
\end{enumerate}
The first three stages are schematically shown in Fig.~\ref{fig:sketch}.
}
  
\bigskip

{We first comment on the latest stage, when the pulsar has already escaped from the SNR's boundaries and propagates through the unshocked ISM. Given that in projection the PSR is within the SNR, a high velocity along the line of sight is needed. This appears to be barely consistent with the recent measurement of the pulsar's line-of-sight velocity \citep[][]{2021NatAs...5..788Y}, although large uncertainties preclude firm conclusions. If the ISM around S147 is cold, one might expect signs of a strong bow shock in ${\rm H_{\alpha}}$ and X-rays ahead of the pulsar. We consider the lack of such signs as an argument against this scenario.}

Three other options include {, respectively,} propagation of the pulsar in the free expanding ejecta, in the medium behind the reverse shock, and in the medium behind the reflected reverse shock, which has overtaken the pulsar after the rebound in the center. For the last two options, the medium might be composed either of the ejecta or shocked ISM gas, and in the more realistic situation, it is likely a mixture of both. What could be the mechanisms capable of producing an almost straight and narrow 2.5 pc-long radio emitting structure?    

The early phase of a PWN evolution within the free expanding ejecta has been considered recently by \citealt{Blondin_Chevalier17}, concluding that the PWN-ejecta interface might be susceptible to instabilities, resulting in strong distortions of the interface and even penetrating filaments left behind the PWN boundary. An important aspect of this situation is, however, the relative smallness of the resulting structures - even taking into account the fact that spin down luminosity of \psr\, should have been much (e.g. a hundred times) higher than the present day one, the size of the PWN is unlikely to exceed a fraction of a pc at most. 

An interesting opportunity, in this case, might be connected to the freely expanding nature of the ejecta flow resembling closely the cosmological Hubble flow for any fluid element. Indeed, every structure connecting two points is getting stretched, given that it evolves passively following the flow streamlines. Hence a straight perturbation e.g. in the form of a small magnetised filament, seeded in a very early stage of the PWN formation might get stretched to a length of a few pc or so, and later filled with the relativistic particles produced by the pulsar over its lifetime. 

In light of this scenario, particularly important is the lower limit on the magnetic field strength obtained from the upper limit on X-ray surface brightness. Indeed, the unshocked free expanding ejecta is expected to possess only very weak magnetic field following $B\propto (R/R_0)^{-2}$, where $R$ is the radial coordinate and $R_0\sim R_{*}$ is the initial radius comparable to the size of the progenitor star. For $R\gtrsim15$ pc and any realistic size of the star, the obtained lower limit on the magnetic field strength inside the filament would translate into way too high magnetic field of the progenitor star.

{
In the "pulsar-in-ejecta" scenario another possibility could be that a magnetized channel in the ejecta was created by an single (early) collimated energy release episode by the pulsar and directed along its rotation axis. However, if this structure has been created recently, a relatively low bulk Lorenz factor ($\Gamma\sim1$) and/or very high degree of collimation ($\theta\sim0.1$ deg) of the outflow is required even assuming that the ejecta gas density is low ($n_{\rm ej}\sim10^{-3}$ cm$^{-3}$) and the integrated outflow energy is as high as 
$E\sim L_{\rm sd} t\sim 10^{47}$ erg (${L_{\rm sd}}/{10^{35}\rm erg/s}) ({t}/{30 \rm kyr}$) in order produce the length $l\sim2.8$~pc $({E}/{10^{47}\rm erg})^{1/3}({n}/{10^{-3} \rm cm^{-3}})^{-1/3} (\Gamma\theta_{\rm deg})^{-2/3}$
\citep[cf. e.g. Eq.~5 in ][]{2002A&A...388L..40H}.

Unless the magnetic field strength in the radio filament is extremely high, allowing particles with $\gamma\ll1000$ being responsible for the observed GHz radio emission, this would imply the presence of some hidden flow dominating the energy and even more the momentum budget. Although such a case of mass loading might be relevant for the jets of accreting black holes and neutron stars (e.g. Cyg X-1 or Cir X-1), such a situation for an isolated pulsar appears very exotic. Creation of this structure at an early stage and then its stretch by a factor of $\delta\sim100$ does not change this argument substantially, as the density of the ejecta was $\delta^2\sim10^4$ times more while the available energy accumulation time $\delta\sim100$ times shorter.
}

In a spherically symmetric case (which of course is a severe oversimplification), the reverse shock propagates inwards, so its interaction with the PWN is not expected to produce a radio-emitting substructure directed away from the SNR's centre \citep[e.g. ][]{2003A&A...397..913V,2017ApJ...844....1K}. Hence, in this scenario, the filament might be associated with a magnetic field substructure present in the shocked ejecta or ISM medium. Since certain mechanisms of the magnetic field generation are conceivable to operate in the shocked ejecta, the required magnetic field strength is not a big problem anymore. This is of course even more true for the shocked ISM case. Given the likely inhomogeneity of the ejecta and ISM, in particular the possible presence of clumps and small clouds capable of shadowing radially elongated "shadows" after the passage of a shock behind them, the radial orientation of the radio filament might look rather natural. 

A potentially serious issue with this model is the transient nature of this phenomenon and the absence of traces from previous such episodes along the track of the pulsar (given that the lifetime of the radio-emitting electrons is longer than the characteristic lifetime of the system). A possible explanation might be that the pulsar is only now entering the region with significantly strong and structured magnetic fields, e.g. corresponding to the compressed magnetic field of the interstellar medium or the fingers produced by the Rayleigh-Taylor (RT) instability \citep{Blondin_Chevalier17}.

{In such a case, the filament might be a radio analogue of the Guitar \citep[]{2007A&A...467.1209H,2022ApJ...939...70D}, Lighthouse  \citep[]{2014A&A...562A.122P,2023ApJ...950..177K}, and PSR~J1509-5850 \citep{2016ApJ...828...70K} and PSR~J2030+4415 \citep{2020ApJ...896L...7D,2022ApJ...928...39D}
X-ray filaments, possibly resulting from the escape of the relativistic particles from the PWN in action \citep{2008A&A...490L...3B,2017SSRv..207..235B,2019MNRAS.485.2041B,2019MNRAS.490.3608O}. In particular, the direction of the extended X-ray filament for PSR~J1509-5850 also appears to be in front of the pulsar's direction of motion \citep{2016ApJ...828...70K}. In contrast to these cases, however, the newly discovered radio-emitting filament might present a case when low-energy relativistic particles are capable of escaping the nebula. Hence, the \psr's radio filament might offer a link between the X-ray filaments and non-thermal filaments observed in the Galactic Center \citep{1984Natur.310..557Y,2022ApJ...925..165H,2022ApJ...925L..18Y}, if the latter are indeed powered by particles escaping from PWNe \citep[e.g.][]{2019MNRAS.489L..28B}.
Relatively low magnetic filed inside the ejecta and/or absence of the converging shocks configuration might be the factor allowing escape and energetic dominance of the lower energy particles in the \psr's case \citep[][]{2017SSRv..207..235B}. 
}

{The scenario of the outwardly propagating {\it reflected} reverse shock that moves faster than the pulsar can explain the direction of the filament by the {most recent episode of the PWN's ram pressure stripping, however leaving open the question of the apparent absence of the tail created by the primary reverse shock.} For a strongly non-spherical geometry (e.g. presence of a dense cloud or pre-existing asymmetric cavity), even a regular reverse shock might move at a large angle to the radius and sweep the tail.  In such a case, this could be an analogue of the tail in Vela X \citep[e.g., ][]{2018ApJ...865...86S}. {In this regard, it is interesting also no note that S147 shows slight asphericity in form of elongation perpendicular to the pulsar's proper motion, as well to the direction of the radio filament \citep[e.g.][]{2006A&A...454..239G}.}  }

{Finally, we summarize how the future (radio) observations will be helpful in revealing the nature of the newly discovered radio filament:
\begin{itemize}
\item spectral index gradient $\rightarrow$ age and $B$
\item polarization in radio $\rightarrow$ morphology of the field
\item surrounding region at lower frequencies
$\rightarrow$ search for "earlier" episodes (relevant only for weak fields)
\item imaging of the starting and ending regions of the filament $\rightarrow$ resolving the region between the pulsar and the filament, as well possible transition/interaction with the H$_\alpha$-bright filament.   
\end{itemize}
}

\section{Conclusions}
\label{s:conclusions}

{We report the discovery of a 2.5-pc-long ({in projection}) and narrow (with aspect ratio $\gtrsim10$) one-sided {radio-emitting (at $\sim {\rm GHz}$ frequencies) filament } tentatively associated with the pulsar PSR~J0538+2817 in the Spaghetti Nebula. Contrary to the known cases of pulsar's radio tail, the filament of PSR~J0538+2817 appears to be directed ahead in the direction of the pulsar's proper motion. While to establish the exact nature of this filament more observations are needed, e.g. providing polarization and spectral index measurements, this object might be a radio analogue of X-ray bright filaments in the Lighthouse and Guitar nebulae, indicating that relatively low energy radio-emitting electrons are able to escape from the PWN. A tantalising connection can also be made with the radio filaments in the Galactic Centre region. We speculate on several scenarios including (i) motion of the PSR together with the freely expanding ejecta threaded by magnetic field lines (young age scenario) and (ii) motion of the  PSR+PWN relative to the shocked gas and an episode of a by-chance reconnection of the bow shock with the ambient magnetic field. These and other scenarios require some strong assumptions, emphasising the unique properties of this object.}

\bigskip

\section*{Acknowledgments}
{We are grateful to B\"arbel Koribalski for a helpful discussion. We thank the papers referee, Dr. Maxim Lyutikov, for useful suggestions.} IK acknowledges support by the COMPLEX project from the European Research Council (ERC) under the European Union’s Horizon 2020 research and innovation program grant agreement ERC-2019-AdG 882679.  AB was supported by the RSF grant 21-72-20020, his simulations were performed at the Joint Supercomputer Center JSCC RAS and at the``Tornado'' subsystem of the Peter the Great Saint-Petersburg Polytechnic University Supercomputing Center. 

This scientific work uses data obtained from Inyarrimanha Ilgari Bundara / the Murchison Radio-astronomy Observatory. We acknowledge the Wajarri Yamaji People as the Traditional Owners and native title holders of the Observatory site. CSIRO’s ASKAP radio telescope is part of the Australia Telescope National Facility (https://ror.org/05qajvd42). Operation of ASKAP is funded by the Australian Government with support from the National Collaborative Research Infrastructure Strategy. ASKAP uses the resources of the Pawsey Supercomputing Research Centre. Establishment of ASKAP, Inyarrimanha Ilgari Bundara, the CSIRO Murchison Radio-astronomy Observatory and the Pawsey Supercomputing Research Centre are initiatives of the Australian Government, with support from the Government of Western Australia and the Science and Industry Endowment Fund. This paper includes archived data obtained through the CSIRO ASKAP Science Data Archive, CASDA (https://data.csiro.au).

This research made use of \texttt{Montage}\footnote{\url{http://montage.ipac.caltech.edu}}. It is funded by the National Science Foundation under Grant Number ACI-1440620, and was previously funded by the National Aeronautics and Space Administration's Earth Science Technology Office, Computation Technologies Project, under Cooperative Agreement Number NCC5-626 between NASA and the California Institute of Technology. Tables manipulations have been performed using the \texttt{TOPCAT/STILTS} software \citep{2005ASPC..347...29T}. We acknowledge the use of data provided by the Centre d'Analyse de Données Etendues (CADE), a service of IRAP-UPS/CNRS \footnote{\url{http://cade.irap.omp.eu}}. This research has made use of the SIMBAD database, operated at CDS, Strasbourg, France. Some of the figures were
produced using the cubehelix color scheme developed by Dave Green \citep[][]{2011BASI...39..289G}.

\section*{Data availability}

All data used in this work is publicly available at astrophysical databases for radio observations (including NVSS, CGPS, ASKAP and Urumqi data), \textit{Chandra} and \textit{XMM-Newton} archives, and IGAPS web page (\url{http://www.star.ucl.ac.uk/IGAPS/}).


\bibliographystyle{mnras}
\bibliography{output} 







\bsp	
\label{lastpage}
\end{document}